\documentclass[english]{iopart}
\usepackage[T1]{fontenc}
\usepackage[latin9]{inputenc}
\usepackage{amssymb}
\usepackage{color}

\usepackage{graphicx}
\newcommand{\myCol}{black}
\makeatletter

\usepackage{iopams}
\usepackage{setstack}

\@ifundefined{showcaptionsetup}{}{
 \PassOptionsToPackage{caption=false}{subfig}}
\usepackage{subfig}
\makeatother

\usepackage{babel}
\begin{document}

\title{Critical point correlations in random gaussian fields}

\author{Avraham Klein and Oded Agam}
\address{The Racah Institute of Physics, The Hebrew University, Jerusalem 91904, Israel}
\ead{avraham.klein@mail.huji.ac.il}
\begin{abstract}
  We consider fluctuations in the distribution of critical points - saddle points, minima and maxima - of random gaussian fields. We calculate the asymptotic limits of the two point correlation function for various critical point densities, for both long and short range. We perform the calculation for any dimension of the field, provide explicit formulae for two and three dimensions, and verify our results with numerical calculations.
\end{abstract}
\maketitle

\section{Introduction}

Random gaussian fields constitute an important paradigm in physics. They serve as a simplified model for diverse physical ingredients such as the energy landscape of glassy systems \cite{Fyodorov2004,Fyodorov2004a,Fyodorov2007a,Bray2007}, wave functions of quantum systems with chaotic classical dynamics  \cite{Berry1977}, and the fluctuations of the Cosmic Microwave Background radiation \cite{Strauss1999}. The critical points of these gaussian fields - minima, maxima and saddle points - characterize the behavior of the entire field and thus carry a large amount of information. For instance, large scale structure in the universe can be analyzed using the critical-point density \cite{Vogel2008}. In other cases critical points are directly related to physical phenomena such as reflection patterns of sunlight from the sea surface \cite{Longuet-Higgins1960}, and the onset of glassy behavior in a complex energy landscape, see \cite{Fyodorov2004} and references therein. An additional novel context  where critical points play an important role is that of cold atoms in disordered optical potentials \cite{Horak1998,Lye2005,Sanchez-Palencia2007}.
In these systems, proper detuning of the light frequency forces the atoms to gravitate to the local extremum points of the intensity, so that statistics of the critical points map directly to statistics of the atom positions.

Most studies of critical points have focused on their density. However, in order to
characterize their statistical properties it is also important to understand their
correlation functions. In particular, in order to calculate fluctuations of the number of critical points, within a given volume of the system, one needs the two-point correlation function.

In this work we shall address the problem of pair correlations of critical points. The one-dimensional version of this problem was solved many years ago \cite{Rice1944,Rice1945}.  In two dimensions, some asymptotic expressions are known \cite{Hohmann2009,Foltin2003}, along with some numerical data \cite{Freund1998} (see also \cite{Dennis2003,Foltin2003a}). {\color{\myCol} For a comprehensive review, concentrating on wave systems, see \cite{Dennis2007} and references therein}. We shall {\color{\myCol} provide a detailed derivation for} the general asymptotic formula of critical point correlations in any dimension $d\geq2$, for both long and short distances. {\color{\myCol} Some of this work, especially in the long-distance limit, may be seen as an expansion and generalization of methods developed previously, see e.g. \cite{Wilkinson2004,Urbina2006,Hohmann2009}}.

There are several types of densities that can be associated with critical points. In order to present them,
let $\phi({\bf r})$ be a gaussian random field  in $d$ spatial dimensions, ${\bf r}= (r_1, r_2 \cdots r_d)$,  with zero mean and an isotopic correlation function
\begin{eqnarray}
\langle \phi({\bf r}) \rangle =0, ~~~~~~~\langle \phi({\bf r})\phi({\bf r}') \rangle= G\left(\left|{\bf r}-{\bf r}' \right|\right),
\end{eqnarray}
We shall assume that the second derivative of $G(r)$ is finite at $r=0$ and vanishes at $r \to \infty$. We note that in general this implies that $r^{-\gamma} G(r) = 0$ vanishes in the limit $r\to \infty$, for $0\leq\gamma<2$.

Critical points are the points where the gradient of the field vanishes, and their density
is given by the well known Kac-Rice formula:
\begin{eqnarray}
\rho_{us}({\bf r})=\left|\det H\right| \delta\left(\boldsymbol{\nabla} \phi\right) \label{eq:usigned-density}
\end{eqnarray}
where $H$ is the Hessian matrix whose elements are the second derivatives of the field:
\begin{eqnarray}
H_{ij}({\bf r})=\frac{\partial^2\phi({\bf r})}{\partial r_i \partial r_j}\label{eq:Hessian}
\end{eqnarray}
This density, which we shall refer to as the \emph{unsigned density} of critical points gives an equal weight and sign to each one of the critical points. There are, however, many other possible choices of density. For instance one may define
the \emph{signed density} as in Eq.~(\ref{eq:usigned-density}) but omitting the absolute value operation.
In two spatial dimensions this density assigns a plus sign to minima and maxima points and a minus sign to saddle points. Another possibility is to define a density of only minima points of the gaussian field. The main focus of this work is the correlation functions of the unsigned density and the density of minima points:
\begin{eqnarray}
{\cal C}(r)=\langle \rho({\bf r}) \rho (0) \rangle  \label{eq:Correlation-def}
\end{eqnarray}
where $\rho({\bf r})$ denotes one of these critical-points densities. However, the methodology we shall develop in order to calculate these functions can be generalized in a straightforward manner to the correlation functions of different types of critical point densities.

The central result of this work is that the asymptotic behavior of the correlation function is
\begin{eqnarray}
{\cal C}(r) = \left\{ \begin{array}{cc} \langle \rho \rangle^2 + \alpha_1 \nabla^{4}G\left( r \right) + \alpha_2 \Tr \left[H_G(r)\right]^2 & r \gg \sigma \\ \alpha_3 r^{2-d+k}  & 0<r \ll \sigma \end{array} \right. \label{eq:main-result}
\end{eqnarray}
where  $\langle \rho \rangle$ is the average density of critical points, $\alpha_1, \alpha_2$ and $\alpha_3$ are constants which depend on the type of the critical-point density and the dimensionality of the system $d$, \begin{equation}
  \label{eq:Hg}
  \left[H_G(r)\right]_{ij}=\frac{\partial^2G({r})}{\partial r_i \partial r_j}
\end{equation}
is the Hessian matrix of the correlation function of the gaussian field, and $\sigma$ is its  typical length scale. In this formula $k=0$ corresponds to the correlation of the unsigned density, while $k=3$ corresponds to the correlation of mimina points. Notice that the correlation function  also contains a $\delta$-function at the origin, ${\cal C}\left(r\to 0\right) = \langle\rho\rangle\delta\left({\bf r}\right)$, due to trivial self correlations of the critical points.

The paper is arranged as follows: In
 Sec. \ref{sec:General-asymptotic-expressions} we shall sketch the derivation of asymptotic form of the correlation function (\ref{eq:Correlation-def}) deferring details to \ref{sec:full-deriv-asympt}. Then we shall apply these results to find the correlation functions in two and three dimensions, in the long-range asymptotic limit. Finally, we will discuss some examples and conclusions in Sec. \ref{sec:discussion}. Appendix A contains the technical details of the derivation, and Appendix B provides information regarding the numerical calculations which were performed in order to compare Eq. (\ref{eq:main-result}) with the exact result.

\section{\label{sec:General-asymptotic-expressions}Asymptotic expressions for  ${\cal C}(r)$}
In this section we shall sketch the calculation of the asymptotic behavior of the pair correlation function  of critical points. We start with a general formalism in which we define the various ingredients of the problem needed for our calculation. Then we show how to obtain the long range and the short range asymptotic behavior of the correlation function of critical points.

\subsection{\label{sub:Presentation-of-the}General formalism}

Consider the critical points of a $d-$dimensional homogenous and isotropic Gaussian random field $\phi\left({\bf r}\right)$, ${\bf r}  \in\mathbb{R}^{d}$. The unsigned critical point density is defined by Eqs.  ~(\ref{eq:Hessian}) and ~(\ref{eq:usigned-density}). Here the Dirac $\delta$-function $\delta\left(\boldsymbol{\nabla} \phi\right)$
selects points with zero gradient, and the Hessian determinant provides the correct normalization since the integral over an infinitesimal region around the critical point is unity.

This density does not distinguish between extrema and saddle points. However, in this work we shall also be interested in the density of the local minima of the field only.
This density may be expressed as
\begin{eqnarray}
\rho_{min}\left({\bf r}\right) & = &
\det H ~\prod_{j=0}^{d-1}\Theta\left(\det H^{\left(j\right)}\right) ~\delta\left(\boldsymbol{\nabla} \phi\right) \label{eq:rhoMin},
\end{eqnarray}
where  $\det H^{\left(j\right)}$ is the $j$-th minor of $H$,
i.e. the determinant of the matrix obtained when the  first $j$ rows and columns of $H$ are removed (the zeroth minor is defined as $H^{(0)} =H$). To understand this result notice that minimum points are associated with the case where all eigenvalues of the Hessian matrix $\lambda_j$ ($j=1,\cdots, d)$) are positive (Note that the Hessian matrix is symmetric and therefore its eigenvalues are real). But this condition can be expressed in terms of determinants of the minors since  $\prod_{j=1}^{d}\Theta\left(\lambda_{j}\right)=
\prod_{j=1}^{d}\Theta\left(\prod_{j=i}^{d}\lambda_{j}\right)
=\prod_{j=0}^{d-1}\Theta\left(\det H^{\left(j\right)}\right)$.

The translational invariance and isotropy of the correlation function of the gaussian field implies that the pair correlation function of critical points depends only on the distance between the two points and therefore one may arbitrarily choose these points to be: one at the origin, ${\bf r}_1=(0,\cdots,0)$, and the other at ${\bf r}_2= (r,0,\cdots,0)$. Then in order to perform the average the product of two densities over the gaussian field it is sufficient to consider the joint probability distribution of the gradients of the fields and the second derivatives at the two previously chosen points ${\bf r_1}$ and ${\bf r}_2$. It is thus convenient to define the vector:
\begin{eqnarray}
\label{eq:Psi-def}
\boldsymbol{\Psi}  =  \left( \boldsymbol{\nabla} \phi ({\bf r}_1),
 \boldsymbol{\psi}_1 , \boldsymbol{\nabla} \phi ({\bf r}_2),  \boldsymbol{\psi}_2) \right),
\end{eqnarray}
where $\boldsymbol{\psi}_\nu$ is a vector constructed from the elements of the Hessian matrix, i.e.
the second derivatives of the gaussian field:
\begin{eqnarray}
\boldsymbol{\psi}_\nu =  \left(\{H_{i,j} ({\bf r}_\nu)\} \right) ~~~~~~ (i\leq j).
\end{eqnarray}

Being expressed in terms of derivatives of a Gaussian field, the vector $\boldsymbol{\Psi}$ is normally distributed, and thus defined by its mean (which vanishes) and the correlation matrix:
 \begin{eqnarray}
B  =  \left\langle \boldsymbol{\Psi}\otimes\boldsymbol{\Psi} \right\rangle \label{B-matrix}
\end{eqnarray}
where $\otimes$ denotes an external product. Thus
\begin{eqnarray}
{\cal C}(r) = \frac{1}{(2\pi)^n\sqrt{ \det B}} \int d \Psi \rho (0) \rho(r) \exp \left( -\frac{1}{2} \boldsymbol{\Psi} B^{-1} \boldsymbol{\Psi} \right) \label{eq:Correlation1}
\end{eqnarray}
where $2n= 2d+ (d-1)d$ is the number of components of the vector $\boldsymbol{\Psi}$, and $d \Psi$ represents
the an infinitesimal volume in the $2n$-dimensional space.

Now the correlation matrix has the general form
 \begin{eqnarray}
B  = \left( \begin{array} {cc} b(0)  & b(r) \\ b^t(r) & b(0) \end{array} \right),
\end{eqnarray}
where the matrix elements of $b(0)$ (which is independent of $r$) correlate components  of $\boldsymbol{\Psi}$ at a single point in space, while those of $b(r)$ (and its transposed matrix  $b^t(r)$) correlate components of the field associated points that are a distance $r$ apart.  This structure is convenient for an asymptotic expansion since  at distance greater than the correlation length of the gaussian field  $b(r) \to 0$ , while in the opposite limit, $r \to 0$, the matrix $b(r)$ converges to  $b(0)$. Thus $b(r)$ and $b(r)-b(0)$, respectively, may be treated perturbatively in order to obtain the asymptotic behavior of ${\cal C}(r)$.

In order to take advantage of this property it is convenient to use the Hubbard-Stratonovich transformation and represent the integral (\ref{eq:Correlation1}) in the form:
 \begin{eqnarray}
{\cal C}(r) = \frac{1}{(2\pi)^{2n}} \int d\Psi \rho (0) \rho(r) \int d \Pi \exp \left(- \frac{1}{2} \boldsymbol{\Pi} B \boldsymbol{\Pi} + i \boldsymbol{\Pi} \cdot \boldsymbol{\Psi} \right)~~~~ \label{eq:Correlation2}.
\end{eqnarray}

The integration over the components of $\boldsymbol{\Psi}$ associated with the field gradients can be readily carried out, and by performing also the integration over their Fourier conjugates we obtain:
\begin{eqnarray}
\!\!\!\!\! \!\!\!\!\!\!\!\!{\cal C}(r) = \frac{1}{(2\pi)^{2n-d}\sqrt{\det A}} \int d\psi \tilde{\rho} (0) \tilde{\rho(r)} \int d \pi \exp \left(- \frac{1}{2} \boldsymbol{\pi} \tilde{B} \boldsymbol{\pi} + i \boldsymbol{\pi} \cdot \boldsymbol{\psi} \right)~~~~~~~~~ \label{eq:Correlation3},
\end{eqnarray}
where $\tilde{\rho}({\bf r}) = \rho({\bf r})/\delta(\boldsymbol{\nabla} \phi({\bf r}))$,  and the remaining integral is over  the vector
\begin{eqnarray}
\boldsymbol{\psi}  =   \left( \boldsymbol{\psi}_1 , \boldsymbol{\psi}_2 \right) =
\left(\boldsymbol{\psi}\left({\bf r}_1\right),\boldsymbol{\psi}\left({\bf r}_2\right)\right),
\end{eqnarray}
and its Fourier conjugates which we denote by $\boldsymbol{\pi}$. The matrices $A$ and $\tilde{B}$ may be expressed in terms of submatrices of $B$. Details on their structure may be found in \ref{sec:full-deriv-asympt}.

\subsection{Long range asymptotics}
\label{sec:long-range-asymptotics}

In order to obtain the asymptotics in this limit we use the fact that the off-diagonal blocks of  $B$ approach $0$ in the limit $r \to \infty$, and therefore may be treated as small parameters.  It is thus convenient to introduce a dummy parameter to $B$ {\color{\myCol}\cite{Wilkinson2004}}:
\begin{equation}
  \label{eq:bEta}
  B  = \left( \begin{array} {cc} b(0)  & \eta  b(r) \\ \eta  b^t(r) & b(0) \end{array} \right)  \nonumber.
\end{equation}
$\eta$ will serve as a bookkeeping parameter of the pertubation theory, and at the end of the calculation will be set to 1. With this addition, $A$ and $\tilde{B}$ appear in the form
  \begin{eqnarray}
  A =  \left( \begin{array}{cc} a(0) & 0 \\ 0 & a(0) \end{array} \right)+ \eta \left( \begin{array}{cc} 0 & a(r) \\  a(r) & 0 \end{array} \right)\label{eq:longA}, \\
  \tilde{B} =  \left( \begin{array}{cc} \tilde{b}(0) & 0 \\ 0 & \tilde{b}(0) \end{array} \right) + \eta \left( \begin{array}{cc} 0 &  \tilde{b}(r) \\ \tilde{b}(r) & 0 \end{array} \right) - \eta^2 \left( \begin{array}{cc} d_1(r) &  d_2(r) \\ d_2(r) & d_1(r) \end{array}\right) \label{eq:B_Def},
  \end{eqnarray}
where $a(r),\tilde{b}(r),d_1(r)$, and $d_2(r)$ are submatrices which approach zero as $r\rightarrow\infty$ (see Appendix A for details). Substituting these matrices in (\ref{eq:Correlation3}) and expanding in $\eta$ give the formal expansion of the correlation function,
\begin{eqnarray}
{\cal C}(r) = {\cal C}_0+ \eta {\cal C}_1(r)+ \eta^2 {\cal C}_2(r)+ \cdots, \label{eq:CorrelationExpansion}
\end{eqnarray}
where
\begin{eqnarray}
  \label{eq:c0}
{\cal C}_0  =\frac{\langle  \tilde{\rho}(0)  \tilde{\rho}(r) \rangle_0}{ (2\pi)^d \det a(0)}  =\left\langle \rho\right\rangle^2 \label{eq:C0}
\end{eqnarray}
is the disconnected part of the correlation function, and here we use the notation
\begin{eqnarray}
\label{eq:isoIntegral}
 \langle \cdots \rangle_0 =\frac{1}{\sqrt{det (2 \pi \tilde{B}_0)}} \int d\psi (\cdots)  \exp\left( -\frac{1}{2} \boldsymbol{\psi}  \tilde{B}_0^{-1}
  \boldsymbol{\psi}\right),
\end{eqnarray}
where $\tilde{B}_0= \tilde{B}|_{\eta=0}$.

The leading order corrections are obtained by expanding the integrand of (\ref{eq:Correlation3}) in $\eta$, and performing the resulting integral. The result takes the form
\begin{eqnarray}
{\cal C}_1(r) = \frac{1}{2} \Tr \left(K F \right) \label{eq:c11},
\end{eqnarray}
where
\begin{eqnarray}
\label{eq:K}
K= \tilde{B}_0^{-1} \left ( \begin{array} {cc} 0  & b(r) \\ b(r) & 0 \end{array} \right)\tilde{B}_0^{-1}
\end{eqnarray}
and
\begin{eqnarray}
\label{eq:F}
F= \frac{1}{ (2\pi)^d \det a(0)}  \langle \tilde{\rho}(0) \tilde{\rho}(r) \boldsymbol{\psi} \boldsymbol{\psi}^t \rangle_0.
\end{eqnarray}
The leading contribution to $\mathcal{C}_2$ depends on the behaviour of $G(r)$ and its derivatives. {\color{\myCol} If, for simplicity's sake, we assume that $G(r)$ {\em does  not have a characteristic wavelength}, i.e is not oscillatory or exponentially decaying, then we have:
\begin{eqnarray}
{\cal C}_2(r) \simeq {\cal C}_0  \frac{1}{2}\Tr\left[ a(0)^{-1} a(r) \right]^2. \label{eq:c12}
\end{eqnarray}
If $G(r)$ has a characteristic wavelength a more complicated expression is necessary, and appears in (\ref{eq:c2-full}) of the Appendix.}

This result gives the leading order behavior of ${\cal C}(r)$ in the long range asymptotic limit. Further details on the structure of $K$, $F$, and the matrices $a(r)$ and $b(r)$, are left for \ref{sec:full-deriv-asympt}.
 For now we just point out that $a(r)=-H_G(r)$, while $\nabla^4 G(r)$ can be constructed from elements of $b(r)$. Thus performing the trace operations in Eqs.~(\ref{eq:c11}), ~(\ref{eq:c12}) leads to our main result, Eq.~(\ref{eq:main-result}), with the constants:
\begin{eqnarray}
  \label{eq:longRangeConsts}
  \alpha_1 & = & {\cal C}_0  f  \frac{d}{d+2} \left[\left. \nabla^4G(r) \right|_{r=0} \right]^{-1}\\
  \alpha_2 & = & {\cal C}_0  d  \left[\Tr H_G(0)^2\right]^{-1}\nonumber
\end{eqnarray}
where $f$ vanishes for the correlation function of the unsigned density, while  it is a constant of order
unity for the minima correlation function. This constant, which depends on the the dimensionality of the system, will be discussed in Sec. \ref{sec:applications} below.

\subsection{\label{sub:Short-range-asymptotics}Short range asymptotics}
As $r \to 0$, the matrix $B$ becomes singular. As a result, both $A$ and $\tilde{B}$ in Eq. ~(\ref{eq:Correlation3}) become singular as well. Therefore, in order to calculate the correlation function, one should handle carefully the divergency  of the prefactor in  Eq. ~(\ref{eq:Correlation3}) in the limit $r \to 0$, and cast $\tilde{B}$ in a form appropriate for expansion. This process is rather technical, and we delegate the body of this discussion to \ref{sec:full-deriv-asympt}, bringing a bare outline here.

We choose $r$ itself to be our expansion parameter. It can then be shown that for $r \to 0$:
\begin{equation}
 \left[\det A\right]^{-1/2} \propto r^{-d} + \cdots \label{eq:detAClose}
\end{equation}
Now, as $r \to 0$, the matrix difference $b(r) - b(0)$ approaches zero.  Therefore, in this limit it is convenient to represent the matrix $\tilde{B}$ in the form:

\begin{eqnarray}
\tilde{B} =  \left( \begin{array}{cc} b_1(r) & b_1(r) \\ b_1(r) & b_1(r) \end{array} \right) + \left( \begin{array}{cc} 0 & \Delta b  \\  \Delta b  & 0 \end{array} \right)
\end{eqnarray}
with $b_1(r) = \tilde{b}(0) - d_1(r), \Delta b = \left[\tilde{b}(r) - \tilde{b}(0)\right] - \left[d_1(r) - d_2(r)\right]$ (see Eq. ~(\ref{eq:B_Def})). This implies $\Delta b \to 0$ when $r \to 0$.  We then perform the unitary rotation
\begin{eqnarray}
\left( \begin{array} {c} \boldsymbol{\psi}_+ \\  \boldsymbol{\psi}_- \end{array} \right)
 = \frac{1}{\sqrt{2}} \left(\begin{array} {cc} 1 & 1 \\  1 & -1 \end{array} \right) \left( \begin{array} {c} \boldsymbol{\psi}_1  \label{eq:rotation}\\ \boldsymbol{\psi}_2 \end{array} \right)
\end{eqnarray}
and a similar rotation to the Fourier conjugate variables $\boldsymbol{\pi}$. This transformation block-diagonalizes the matrix $\tilde{B}$, so that ${\cal C}(r)$ takes the form (\ref{eq:Correlation3}) but with the matrix $\tilde{B}$ replaced by
\begin{eqnarray}
\tilde{B} \to \left( \begin{array}{cc} b_1 + \Delta b & 0 \\ 0 & - \Delta b \end{array} \right)
\end{eqnarray}
An additional transformation block-diagonalizes $b_1 + \Delta b$, and allows one to cast the resulting matrix as a power series in $r$ and perform the integral. The result is
\begin{equation}
  \label{eq:shortIntegral}
 \langle \tilde{\rho}(0) \tilde{\rho}(r)\rangle \propto \left\{ \begin{array}{cc} r^2 & \mbox{unsigned density } \\ r^5 & \mbox{density of minima} \end{array}, \right.~~~~ r \ll \sigma.
\end{equation}
Combining Eqs. ~(\ref{eq:Correlation3}), ~(\ref{eq:detAClose}) and ~(\ref{eq:shortIntegral}) leads immediately to our main result, Eq. ~(\ref{eq:main-result}).

\section{Application in two and three dimensions}
\label{sec:applications}
In this section we shall find an asymptotic expression for the correlation in two and three dimensions. We begin by calculating ${\cal C}_0$ and $F$ given by Eqs. (\ref{eq:C0}) and (\ref{eq:F}) respectively. Both formulae no longer possess an explicit dependence on $r$. As a result, the isotropy of the field $\phi$ greatly simplifies calculations. Thus, finding ${\cal C}_0$ clearly reduces to simply finding the density expectation:
\begin{equation}
  \label{eq:c0-separated}
  {\cal C}_0 \propto \langle\tilde{ \rho}(0)\tilde{\rho}(r)\rangle_0 = \langle\tilde{\rho}(0)\rangle_0^2.
\end{equation}

Isotropy also implies that all non-zero elements of $F$ appearing in ${\cal C}_1$ have the same value $f_1$, thus:
\begin{eqnarray}
  \label{eq:fVal}
  f_1 = {\cal C}_0 f  &\propto& \left(\left\langle \left. \tilde{\rho}(0) \frac{\partial^2\phi(\mathbf{r})}{\partial r_i^2}\right|_{\mathbf{r}=0} \right\rangle_0\right)^2  \nonumber \\
 & =&  \frac{1}{d^2}\left(\langle \tilde{\rho}(0) \Tr H(0)\rangle_0\right)^2
\end{eqnarray}
where $f$ is the constant which has been introduced in Eq. ~(\ref{eq:longRangeConsts}), and $F_{1,n-d+1}$ is an example of one of the relevant non-zero elements of $F$.

The density $\langle\rho\rangle$ (and thus ${\cal C}_0$) has been calculated by others \cite{Longuet-Higgins1960,Vogel2008}, therefore, it remains to find $f$. We devote the rest of this  section  to this calculation, and to be concrete we choose the  correlation function of the gaussian field to be
\begin{equation}
  \label{eq:Gd}
  G_{d}(r) = \left( \frac{1 }{2 r}\right)^{\frac{d -2}{2}} J_{\frac{d -2}{2}} \left(\pi r\right),
\end{equation}
where $d=2,3$ is the dimensionality and $J_\nu$ is the Bessel function of order $\nu$. This form of correlation appears naturally in the context of ballistic chaotic wave systems, the so-called Random Wave Model \cite{Berry1977,Dennis2007}. This choice does not affect the calculation of $f$ but is useful for comparison with numerical results.

\subsection{Two dimensions, $d=2$}
\label{sec:app-two-dimensions}

In two dimensions $\boldsymbol{\psi}=\left(\phi_{xx},\phi_{xy},\phi_{yy}\right)$,
where for brevity we use $x,y$ subscripts to denote derivatives with respect to the corresponding coordinates. In this case we have defined $\boldsymbol{\psi} = \boldsymbol{\psi}_1$ since $\boldsymbol{\psi}_2$ no longer appears in expressions (\ref{eq:c0-separated}) and (\ref{eq:fVal}). By
transforming to the normalized variables \cite{Longuet-Higgins1960,Foltin2003b},
\begin{eqnarray*}
X & \propto & \phi_{xy}\\
Y & \propto & \phi_{xx}-\phi_{yy}\\
Z & \propto & \phi_{xx}+\phi_{yy}
\end{eqnarray*}
one obtains $\det H\propto2Z^{2}-X^{2}-Y^{2},\,\Tr H\propto Z$,
and the Gaussian measure becomes diagonalized. It is now possible to calculate $\left\langle \tilde{\rho}(0)\Tr H(0)\right\rangle_0 $ by transforming to spherical coordinates. A similar derivation is
found in \cite{Foltin2003} and we omit the details.
The final result is $f=0$ for the correlation function of unsigned density. (This is in fact true for any dimension and can be seen from symmetry considerations.) For the minima-minima correlation,
\begin{equation}
  \label{eq:min2dC1}
  f = \frac{64}{27\pi}.
\end{equation}

\begin{figure}[h]
\begin{centering}
\subfloat[\emph{Minima-minima correlation function}]{\begin{centering}
\includegraphics[clip,width=0.8\textwidth]{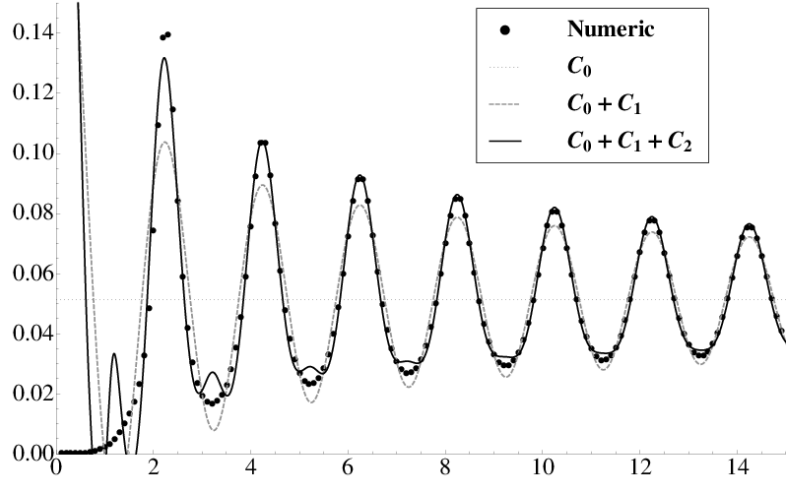}
\par\end{centering}}
\par\end{centering}

\begin{centering}
\subfloat[\emph{Unsigned correlation function}]{\begin{centering}
\includegraphics[clip,width=0.8\textwidth]{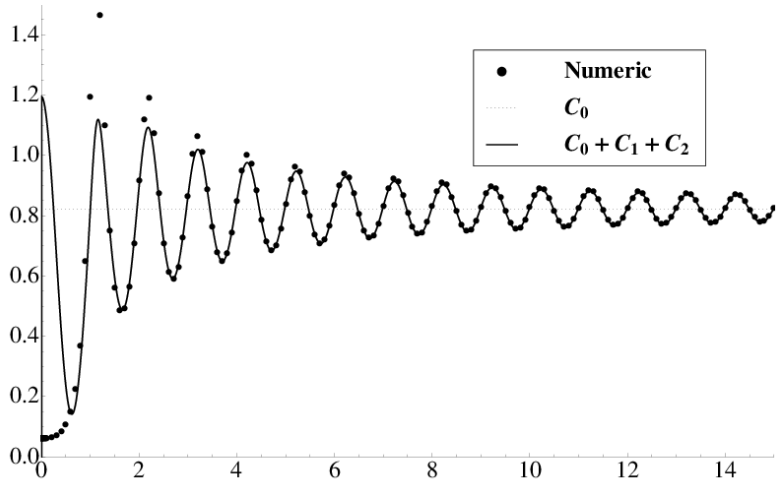}
\par\end{centering}}
\par\end{centering}

\caption{\label{fig:2dFar} The critical point correlation functions for a
gaussian random field in two dimensions, with Bessel  function correlation $G\left(r\right)=J_0\left(\pi r\right)$.
The term ${\cal C}_0$ represents the disconnect part of the correlation function, while
${\cal C}_1$ and  ${\cal C}_2$ are the leading terms of the long range asymptotic expansion. 
The black dots represent the exact result obtained by numerical calculation explained in Appendix B.}
\end{figure}

We conclude this section by referring to figure \ref{fig:2dFar}. This shows a comparison of our analytical asymptotic expression, and a numerical evaluation. Details of the numerical procedure used in order to get this figure can be found in Appendix B.

\subsection{Three dimensions, $d=3$}
\label{sec:three-dimensions}

The critical-point density was found for $d=3$ in the context of the Cosmic Microwave Background radiation \cite{Vogel2008}. In this case there are six variables in the Hessian: $\boldsymbol{\psi} = (\phi_{xx},\phi_{xy},\phi_{xz},\phi_{yy},\phi_{yz},\phi_{zz})$, where as in the two-dimensional case, we write for brevity $\boldsymbol{\psi}=\boldsymbol{\psi}_1$ and the subscripts denote partial derivatives. Here we  transform the Gaussian integral over $\boldsymbol{\psi}$ into an integral over the eigenvalues of the Hessian matrix $H$. To this end it is convenient to represent Gaussian exponent in the form \cite{Fyodorov2004,Vogel2008}:
\begin{eqnarray*}
\boldsymbol{\psi} \left[b(0)\right]^{-1}\boldsymbol{\psi} & = & \frac{1}{4\left\langle \phi_{xy}^{2}\right\rangle }\left(\Tr H^{2}-\frac{1}{d+2}{\Tr}^{2}H\right)
\end{eqnarray*}
 where here we have used the block form of $\tilde{B}_0$ in Eq. (\ref{eq:B_Def}). Employing the standard approach of random matrix theory \cite{Mehta1991} one may transform from the integration variables $\boldsymbol{\psi}$ to to an integration over eigenvalues of $H$, $\boldsymbol{\lambda}=\left(\lambda_{1},\lambda_{2},\lambda_{3}\right)$. The measure of the integral in this case is
\begin{eqnarray}
 d\psi \exp \left( \boldsymbol{\psi} \left[b(0)\right]^{-1}\boldsymbol{\psi} \right)  = \label{eq:eigPDF} \\
 \frac{ \prod_{j=1}^3 d\lambda_j} { 8 \sqrt{2} \pi \langle \phi_{xy}^{2}\rangle^6} \left(\lambda_{2}-\lambda_{1}\right)\left(\lambda_{3}-\lambda_{2}\right)\left(\lambda_{3}-\lambda_{1}\right)\exp\left[-\frac{1}{4A^{2}}\left(\boldsymbol{\lambda} M\boldsymbol{\lambda}\right)\right] \nonumber,
\end{eqnarray}
with  $M_{ij}=\delta_{ij}-\frac{1}{5}$.  Following Ref. \cite{Vogel2008},
%\footnote{Note a slight numerical error in \cite{Vogel2008}, namely a factor $\frac{1}{2}$ in the exponent in Eq. 9 of that paper, as opposed to the correct factor $\frac{1}{4}$ as appearing in Eq. ~(\ref{eq:eigPDF})of this work, leads to different results from ours.%}
further diagonalization of the Gaussian exponent followed by a change to spherical coordinates allows one to calculate  $\langle\tilde{\rho}(0)\Tr H(0)\rangle$ and to deduce the value of the constant $f$ from Eq.
(\ref{eq:fVal}). The result is $f=0$ for the correlation function of unsigned density, while
for the minima-minima correlation:
\begin{equation}
  \label{eq:fMin3}
  f = \frac{31250}{9 (29 - 6 \sqrt{6})^2 \pi}
\end{equation}

\begin{figure}[h]
\begin{centering}
  \subfloat[\emph{Minima-minima correlation function}]{\includegraphics[clip,width=0.8\textwidth]{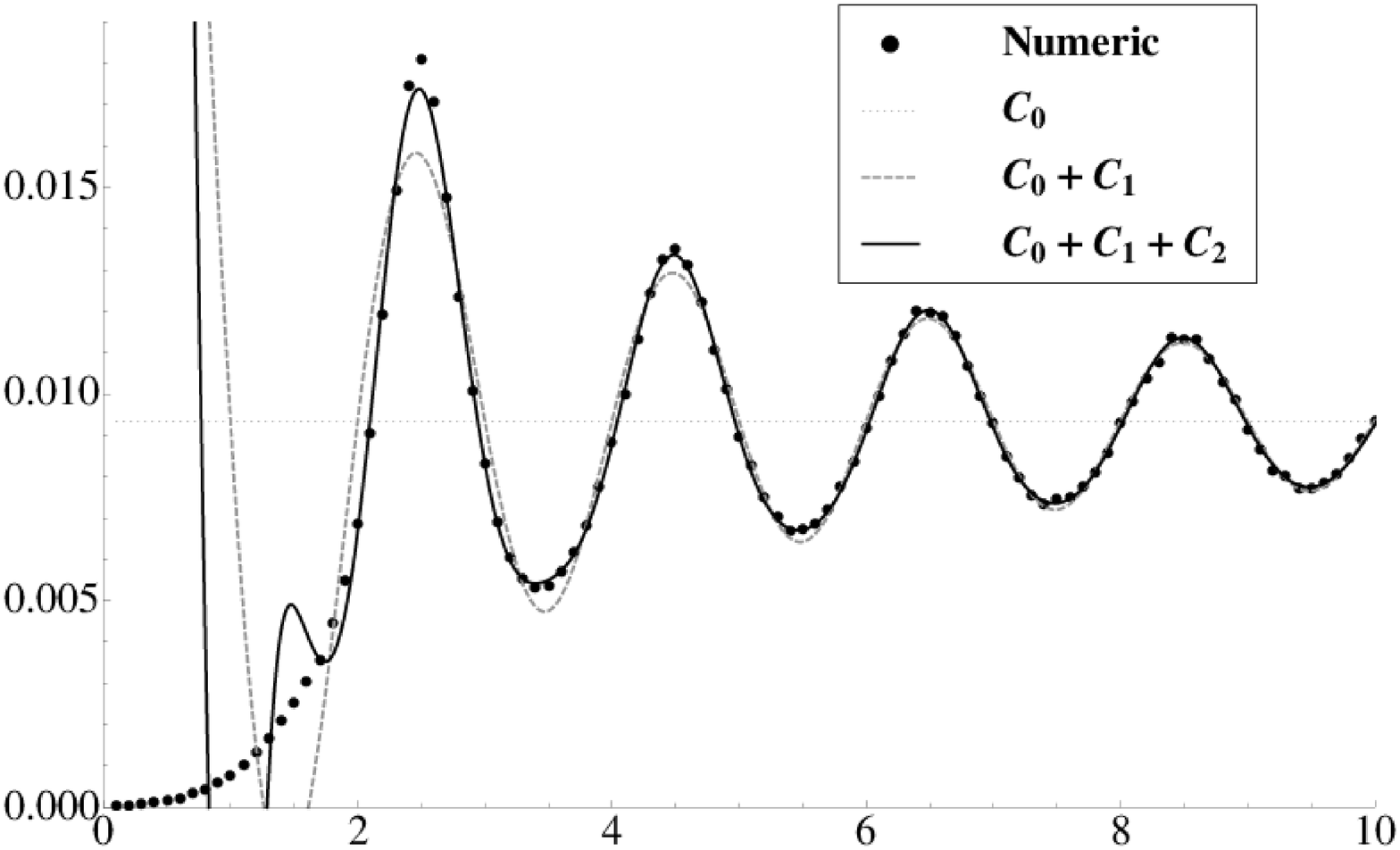}}
\end{centering}

\begin{centering}
\subfloat[\emph{Unsigned correlation function}]{\includegraphics[clip,width=0.8\textwidth]{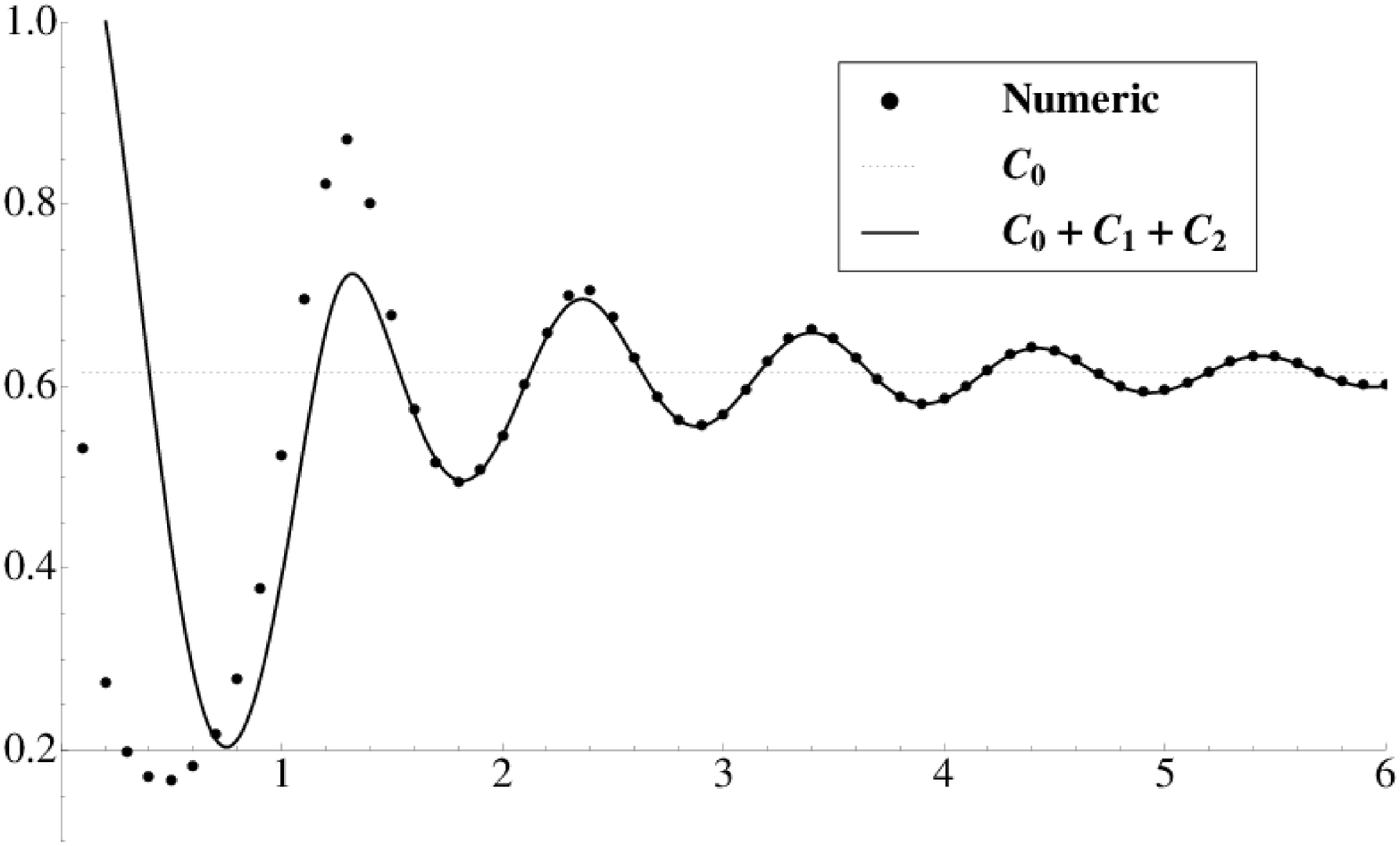}}
\end{centering}

\caption{\label{fig:3dFar}The critical point correlation functions for a
gaussian random field in three dimensions, with $G\left(r\right)=\frac{\sin\pi r}{\pi r}$.
The meaning of lines and the dots in this figure is the same as in Fig.~1} 
\end{figure}

We conclude this section by referring to figure \ref{fig:3dFar}, which shows a comparison of our analytical asymptotic expression, and a numerical evaluation.

\section{Discussion}
\label{sec:discussion}
In this work we calculated the asymptotic behavior of the pair correlation function of critical points, ${\cal C}(r)$ in two limiting regions, $r \to \infty$, and $r \to 0$. We have shown that in both limits the behavior correlation function depends on the types of densities of critical points, i.e. whether it is the unsigned
 density or the the minima-minima density. In the short range asymptotic it is manifested through the power law dependence of the correlation function: ${\cal C}(r) \propto r^{2-d}$ for  of the unsigned correlation and ${\cal C}(r) \propto r^{5-d}$ for minima-minima correlations. 

In the long range asymptotics we have seen that there are two terms contributing to the correlation functions, see Eq.~(\ref{eq:main-result}). However, for the unsigned correlation the prefactor of the term proportional to $\nabla^4 G(r)$ vanishes while for the minima-minima correlation both terms contribute.

 Let us consider some specific examples in order to clarify the nature of the contributions of the two terms, $\nabla^4 G(r)$ and $\Tr \left[H_G(r)\right]^2$, to the long range asymptotic limit.  Consider the case where the correlation function of the gaussian field is an oscillatory function of the form
\begin{eqnarray}
\lim_{r \to \infty} G(r) \longrightarrow A \frac{  \cos (r/\sigma)}{ r^\gamma}\label{eq:G1}
\end{eqnarray}
where $A$ is the amplitude, $\sigma$ is a constant which sets the length scale of the oscillations, while $\gamma >0$ determines the behavior of the envelope of the oscillations. This type of correlation describes, e.g., a field generated by wave chaos \cite{Dennis2007}.
For this case in the asymptotic limit $ r \to \infty$
\begin{eqnarray}
\nabla^4 G(r)  \sim  \frac{A}{\sigma^4}\frac{\cos (r/\sigma) }{r^\gamma} \label{eq:c1G1}
\end{eqnarray}
while
\begin{eqnarray}
\Tr \left[H_G(r)\right]^2  \sim \frac{A^2}{\sigma^4}\frac{\cos^2 (r/\sigma) }{r^{2\gamma}}
\end{eqnarray}
Thus for large enough $r$ the first term, Eq. (\ref{eq:c1G1}), dominate the behavior (unless its prefactor $\alpha_1$ vanishes as for the unsigned correlator).

Consider the case where at large distances the correlation function is described by a simple power law behavior, i.e.
\begin{eqnarray}
  \label{eq:G2}
  \lim_{r \to \infty} G(r) \longrightarrow  \frac{ A }{ r^\gamma}
\end{eqnarray}
 where $\gamma >-2$. This type of correlation, in particular for $\gamma < 0$, is believed to describe the potential landscape of glassy systems \cite{Fyodorov2007b}. For this case we find that
\begin{eqnarray}
  \label{eq:c1G2}
  \nabla^4 G(r)  \sim \gamma (2+ \gamma) (d-\gamma-2)(d-\gamma-4) \frac{ A}{r^{\gamma+4}},
\end{eqnarray}
while
\begin{eqnarray}
  \label{eq:c2G2}
  \Tr \left[H_G(r)\right]^2  \sim  \gamma^2 (\gamma+ d)^2\frac {A^2}{r^{2\gamma+4}}.
\end{eqnarray}
Thus the term (\ref{eq:c1G2}) is dominant for $\gamma > 0$, while (\ref{eq:c2G2}) is dominant for $\gamma < 0$.

Finally, for logarithmic correlation:
\begin{eqnarray}
  \label{eq:G3}
  \lim_{r \to \infty} G(r) \longrightarrow  A \log (r)
\end{eqnarray}
 we have,
\begin{eqnarray}
  \label{eq:d1G2}
  \nabla^4 G(r)  \sim 2(2-d)(d-4) \frac{A}{r^{4}}
\end{eqnarray}
while
\begin{eqnarray}
  \label{eq:d2G2}
  \Tr \left[H_G(r)\right]^2  \sim d \frac {A^2}{r^{4}}
\end{eqnarray}

These examples demonstrate that the asymptotic behavior of the correlations of critical points depend on the type of density, but are also sensitive to the particular value of $\gamma$ and the dimensionality $d$. For instance, for power law correlations with $\gamma=1$, the slower decaying term (\ref{eq:c1G2}) vanishes at $d=3$, while for the logarithmic correlation (\ref{eq:d1G2}) vanishes at $d=2$.

\ack
We would like to thank Y. Fyodorov for very useful discussions, and D. Klein for his advice on the numerical calculations in this work. This research has been supported by the  United States-Israel
Binational Science Foundation (BSF) grant No. 2008278.

\appendix

\section{Details of the derivation in Sec. 2}
\label{sec:full-deriv-asympt}

In this section we provide the details of the derivation of the results brought in Sec. \ref{sec:General-asymptotic-expressions}. To begin with, we present the form of the matrices introduced in Sec. \ref{sub:Presentation-of-the}, Eqs. (\ref{B-matrix}) and (\ref{eq:Correlation3}).
As a first step, redefine the gaussian vector from Eq. (\ref{eq:Psi-def}):
\begin{equation}
  \label{eq:psi-b-definition}
  \boldsymbol{\Psi} = \left( \boldsymbol{\nabla}\phi(\mathbf{r}_1), \boldsymbol{\nabla}\phi(\mathbf{r}_2), \boldsymbol{\psi}_1, \boldsymbol{\psi}_2 \right).
\end{equation}
This reordering of elements is more convenient in terms of the algebra to be carried out shortly. Then we have:
\begin{eqnarray}
  \label{eq:b-full-def}
  B  & = & \langle\boldsymbol{\Psi}\otimes\boldsymbol{\Psi}\rangle  =
  \left(\begin{array}{cc}
    A & M \\
    M^t & \beta
  \end{array}\right) \\
& = & \left(\begin{array}{cc}
    \left(\begin{array}{cc}
        a(0) & a(r) \\
        a(r) & a(0)
      \end{array}\right) & \left(\begin{array}{cc}
        0 & m(r)\\
        -m(r) & 0
      \end{array}\right)\\
    \left(\begin{array}{cc}
        0 & -m(r)^{t}\\
        m(r)^{t} & 0
      \end{array}\right) & \left(\begin{array}{cc}
        b(0) & b(r)\\
        b(r) & b(0)
      \end{array}\right)
  \end{array}\right),
\end{eqnarray}
with the definitions
\begin{eqnarray}
  a(r)_{ij} & = & -\left. \frac{\partial^2 G(\left|\mathbf{r}\right|)}{\partial r_i \partial r_j} \right|_{\mathbf{r}=(r,0\cdots,0)},\\
  b(r)_{ij,kl} & = &\left. \frac{\partial^4 G(|\mathbf{r}|)}{\partial r_i \partial r_j \partial r_k \partial r_l} \right|_{\mathbf{r}=(r,0\cdots,0)},\\
  m(r)_{i,jk} & = & \left. \frac{\partial^3 G(|\mathbf{r}|)}{\partial r_i \partial r_j \partial r_k} \right|_{\mathbf{r}=(r,0\cdots,0)}.
\end{eqnarray}
Here, $\mathbf{r} = \mathbf{r}_2 - \mathbf{r}_1$, and $b$ and $m$ have double indices to denote elements of $\boldsymbol{\psi}_i$. To make things clearer, we show some of these matrices explicitly for the three-dimensional case:
\begin{eqnarray*}
a(0) & = &
  I_d \left|G^{\left(2\right)}\left(0\right)\right|,
\end{eqnarray*}
where $I_d$ is the three dimensional identity matrix, and
\begin{eqnarray*}
  b(0) & = &
  \left(\begin{array}{cccccc}
      3 & 0 & 0 & 1 & 0 & 1\\
      0 & 1 & 0 & 0 & 0 & 0 \\
      0 & 0 & 1 & 0 & 0 & 0\\
      1 & 0 & 0 & 3 & 0 & 1\\
      0 & 0 & 0 & 0 & 1 & 0 \\
      1 & 0 & 0 & 1 & 0 & 3
    \end{array}
  \right)\frac{G^{\left(4\right)}\left(0\right)}{3},
\\
m & = &
\left(\begin{array}{cccccc}
    p & 0 & 0 & q & 0 & q\\
    0 & q & 0 & 0 & 0 & 0\\
    0 & 0 & q & 0 & 0 & 0
  \end{array}
\right);
\begin{array}{l}
  p = \left. \frac{\partial^{3}G(\mathbf{r})}{\partial r_1^3} \right|_{\mathbf{r}=(r,0 \cdots,0)}\\
  q = \left. \frac{\partial^{3}G(\mathbf{r})}{\partial r_1 \partial r_2^2} \right|_{\mathbf{r}=(r,0\cdots,0)}\\
\end{array}
\end{eqnarray*}
It can also be verified that $a(r)$ and $b(r)$ have non-zero elements exactly for the same indices as $a(0)$ and $b(0)$. We plug these definitions into Eq. (\ref{eq:Correlation2}) and perform the integration over the field gradients and their Fourier conjugates. This leads to Eq. (\ref{eq:Correlation3}) with
\begin{eqnarray}
  \label{eq:tildeB-def}
  \tilde{B} & = &\beta - D \\
  D & = & M^t A^{-1} M = \left(
    \begin{array}{cc}
     d_1(r) & d_2(r) \\
     d_2(r) & d_1(r) \\
    \end{array}\right)
\end{eqnarray}
where
\begin{eqnarray}
  d_i(r) & = & \left[a(0)^2-a(r)^2\right]^{-1}(-1)^{i-1}a(s) \label{eq:d_i}\\
  s & = & \left\{\begin{array}{cc} 0 & i = 1 \\ r & i = 2 \end{array}\right.\nonumber
\end{eqnarray}

\noindent

\subsection*{Long range asymptotics}
\label{sec:long-range-appen}
In order to obtain the long range asymptotics to second order in $\eta$ (see Eq.~(\ref{eq:CorrelationExpansion})), we expand the prefactor and gaussian exponent in Eq. (\ref{eq:Correlation3}):
\begin{eqnarray*}
\fl \exp\left[-\frac{1}{2}\boldsymbol{\pi}\tilde{B}\boldsymbol{\pi}\right] &=&  e^{-\frac{1}{2}\boldsymbol{\pi}\tilde{B}_0\boldsymbol{\pi}}\left[1-\eta \cdot \frac{1}{2} \boldsymbol{\pi} \Delta\tilde{B}\boldsymbol{\pi}+\frac{1}{2}\eta^2\left(\frac{1}{2}\boldsymbol{\pi} \Delta\tilde{B}\boldsymbol{\pi}\right)^{2} - \eta^2 \frac{1}{2} \boldsymbol{\pi} D \boldsymbol{\pi}+\ldots\right]\\\fl \det A^{-1/2} & = & \left\{ \left(\det a(0)\right)^{2}\cdot\det\left[I-\left(a(0)^{-1}a(r)\right)^{2}\right]\right\} ^{-1/2}\\
\fl & = &\frac{1}{\det a(0)}\cdot\left[1+\frac{1}{2}\tr \left(a(0)^{-1}a(r)\right)^{2}-\ldots\right],
\end{eqnarray*}
where here we have:
\begin{eqnarray*}
  \Delta\tilde{B} & = &
  \left(\begin{array}{cc}
    0 & b(r) \\
    b(r) & 0
  \end{array}\right) \\
  D & \simeq &
  \left(\begin{array}{cc}
      d_1(r) & 0 \\ 0 & d_1(r)
    \end{array}\right)
\end{eqnarray*}
since at the long range limit $d_2 \ll d_1$, as follows from Eq. (\ref{eq:d_i}). Then we obtain ${\cal C}_1$ from Eq. (\ref{eq:c11}) and ${\cal C}_2$ from Eq. (\ref{eq:c12}), which we can now write in full:
\begin{eqnarray}
  \label{eq:c2-full}
  \fl  {\cal C}_2(r)  = & &\frac{1}{2}\left[{\cal C}_0\Tr\left[ a(0)^{-1} a(r) \right]^2 - {\cal C}_0 \Tr (\tilde{B}_0 L) + \langle\tilde{\rho}\rangle^{-1} \Tr (F L)\right] \nonumber\\
  &  + & \frac{1}{4}\left[ {\cal C}_0 \Tr (\tilde{B}_0 K \tilde{B}_0 K) - 2 \Tr (F K \tilde{B}_0 K) + {\cal C}_0^{-1}\Tr (F K F K)\right]
\end{eqnarray}
with $L = \tilde{B}_0^{-1} D \tilde{B}_0^{-1}$ and $K,F$ as in Eqs. (\ref{eq:K}) and (\ref{eq:F}). {\color{\myCol} We note however that if $G(r)$ diverges at the limit $r\to\infty$, then the $\det A^{-1/2}$ prefactor must be expanded to order $\eta^4$ so as to get the proper subleading behaviour of $\mathcal{C}(r)$.}

Finally, let us examine $F$. We can write it as:
\begin{equation}
  \label{eq:f-def}
  F \propto
  \left(
    \begin{array}{cc}
      \langle\tilde{\rho}(\mathbf{r}_1)\boldsymbol{\psi}_1\boldsymbol{\psi}_1\rangle_{1} \langle\tilde{\rho}(\mathbf{r}_2)\rangle_2 &
      \langle\tilde{\rho}(\mathbf{r}_1)\boldsymbol{\psi}_1\rangle_{1} \langle\tilde{\rho}(\mathbf{r}_2)\boldsymbol{\psi}_2\rangle_2 \\
      \langle\tilde{\rho}(\mathbf{r}_1)\boldsymbol{\psi}_1\rangle_{1} \langle\tilde{\rho}(\mathbf{r}_2)\boldsymbol{\psi}_2\rangle_2 &
      \langle\tilde{\rho}(\mathbf{r}_1)\rangle_{1} \langle\tilde{\rho}(\mathbf{r}_2)\boldsymbol{\psi}_2\boldsymbol{\psi}_2\rangle_2
    \end{array}
  \right)
\end{equation}
where
\begin{equation}
  \label{eq:b0-int}
  \langle\cdots\rangle_i = \frac{1}{\sqrt{\det\left[2\pi b(0)\right]}} \int d\psi_i \left(\cdots\right) exp\left[-\frac{1}{2}\boldsymbol{\psi}_i  b(0)^{-1} \boldsymbol{\psi}_i\right]
\end{equation}
Now, recall that $\tilde{\rho} \propto \det H$. From symmetry properties of the determinant, one can verify that the only non-zero in $\langle\tilde{\rho}\boldsymbol{\psi}\rangle$ (the position $\mathbf{r}_i$ no longer matters) is
\begin{equation*}
  \left\langle \tilde{\rho} \frac{\partial^2 \phi}{\partial r_i^2} \right\rangle = \frac{1}{d} \left\langle \tilde{\rho} \Tr H \right\rangle,
\end{equation*}
and the only non-zeros in  $\langle\tilde{\rho}\boldsymbol{\psi}\boldsymbol{\psi}\rangle$ are ($i\neq j$):
\begin{equation*}
  \left\langle \tilde{\rho} \left(\frac{\partial^2 \phi}{\partial r_i^2}\right)^2 \right\rangle,
  \left\langle \tilde{\rho} \frac{\partial^2 \phi}{\partial r_i^2}  \frac{\partial^2 \phi}{\partial r_j^2} \right\rangle ~\mbox{and}~
  \left\langle \tilde{\rho} \left(\frac{\partial^2 \phi}{\partial r_i\partial r_j}\right)^2 \right\rangle.
\end{equation*}

\subsection*{Short range asymptotics}
\label{sec:short-range-append}

To obtain the short range asymptotics, we assume $G(r)$ is an even function of $r$ and expand it to sixth order:
\[
G\left(r\right)=g_0-\frac{1}{2!}g_2 r^{2}+\frac{1}{4!}g_4 r^{4}-\frac{1}{6!}g_6 r^{6}+\ldots
\]
where $-g_2,g_4$, and $-g_6$ are respectively the second, fourth and sixth derivatives of $G\left(0\right)$. Next, we need to find the prefactor in Eq. (\ref{eq:Correlation3}) and transform $\tilde{B}$ to an appropriate form. To find the prefactor we expand $\det A$ at the limit $r \ll 1$. This yields:
\begin{eqnarray*}
\det A & = & \prod_{i=1}^{d}\left(a(0)_{ii}^{2}-a(r)_{ii}^{2}\right)+\cdots=\frac{\left(g_2g_4r^{2}\right)^{d}}{3^{d-1}}+\cdots
\end{eqnarray*}

After performing the rotation (\ref{eq:rotation}), an explicit calculation gives for three dimensions:
\begin{eqnarray*}
\fl \Delta b &=&\left(\begin{array}{cccccc}
-\frac{g_4^{2}}{2g_2}+\frac{g_6}{2} & 0 & 0 & -\frac{g_4^{2}}{6g_2}+\frac{g_6}{10} & 0 & -\frac{g_4^{2}}{6g_2}+\frac{g_6}{10}\\
0  & -\frac{g_4^{2}}{18g_2}+\frac{g_6}{10} & 0 & 0 & 0 & 0\\
0  &  0  & -\frac{g_4^{2}}{18g_2}+\frac{g_6}{10} & 0 & 0 & 0\\
 -\frac{g_4^{2}}{6g_2}+\frac{g_6}{10} & 0 & 0  & -\frac{g_4^{2}}{18 g_2}+\frac{g_6}{10} & 0 &  -\frac{g_4^{2}}{18 g_2}+\frac{g_6}{10}\\
 0 & 0 & 0 & 0  & \frac{g_6}{10} & 0 \\
-\frac{g_4^{2}}{6g_2}+\frac{g_6}{10} & 0 & 0 & -\frac{g_4^{2}}{18g_2}+\frac{g_6}{10} & 0 & -\frac{g_4^{2}}{18g_2}+\frac{g_6}{10}
\end{array}\right)~ r^{2}+\Or\left(r^{4}\right)\\
\fl b_{0} & = & \left(\begin{array}{cccccc}
0 &  0 & 0 & \frac{g_6r^{2}}{45} & 0 & \frac{g_6r^{2}}{45}\\
0 & 0 &0 &0 &0 &0 \\
 0 & 0 & 0 &0 &0 &0 \\
\frac{g_6r^{2}}{45} & 0 & 0 & \frac{16g_4}{9}-\frac{2g_6r^{2}}{27} & 0 & \frac{4g_4}{9}-\frac{2g_6r^{2}}{27}\\
0 & 0 & 0 & 0 & \frac{2g_4}{3}-\frac{g_6r^{2}}{10} & 0 \\
\frac{g_6r^{2}}{45} & 0 & 0 & \frac{4g_4}{9}-\frac{2g_6r^{2}}{27} & 0 & \frac{16g_4}{9}-\frac{2g_6r^{2}}{27}
\end{array}\right)+\Or\left(r^{4}\right)
\end{eqnarray*}
where $b_0 = b_1 + \Delta b$. It is now useful to diagonalize $b_0$. Let us rewrite the vector $\boldsymbol{\psi}_{+}$, as defined in Eq.~(\ref{eq:rotation}), in the form
\begin{equation*}
  \boldsymbol{\psi}_{+}  =  \boldsymbol{\psi}_{\parallel}+\boldsymbol{\psi}_{\perp},
\end{equation*}
where
\begin{eqnarray*}
\boldsymbol{\psi}_{\parallel} & = & \left(\Phi_{11},\Phi_{12},\dots,\Phi_{1d},0,\dots\right),\\
\boldsymbol{\psi}_{\perp} & = & \left(0,\dots,\Phi_{22},\Phi_{23},\dots,\Phi_{dd}\right).
\end{eqnarray*}
Here $\boldsymbol{\psi}_{\parallel}$ holds the first $d$ elements of $\boldsymbol{\psi}_+$, or in other words all those elements involving derivatives along the line connecting $\mathbf{r}_1$ and $\mathbf{r}_2$. $\boldsymbol{\psi}_{\perp}$ holds all the other elements namely those involving derivatives in direction transverse to the vector $\mathbf{r}_2-\mathbf{r}_1$. We then write the cross correlation matrix of $\boldsymbol{\psi}_+$  in the form
\begin{equation*}
  b_0 = \left(\begin{array}{cc}
c_0 & c_1\\
c_1^{t} & c_2
\end{array}\right)
\end{equation*}
 where $c_0 = \langle\boldsymbol{\psi}_{\parallel}\otimes\boldsymbol{\psi}_{\parallel}\rangle,c_1 = \langle\boldsymbol{\psi}_{\parallel}\otimes\boldsymbol{\psi}_{\perp}\rangle$, and $ c_2 = \langle\boldsymbol{\psi}_{\perp}\otimes\boldsymbol{\psi}_{\perp}\rangle$. This matrix can now be block-diagonalized:
\begin{eqnarray*}
  b_0 & = &
  u^{t}
  \left(
    \begin{array}{cc}
      c_0-c_1c_2^{-1}c_1^{t} & 0\\
      0 & c_2
    \end{array}
  \right) u  =
  u^t \left(\begin{array}{cc}
    c_3 & 0\\
    0 & c_2
  \end{array}\right)u
\end{eqnarray*}
with
\begin{eqnarray*}
  u &= &
  \left(
    \begin{array}{cc}
      I & 0\\
      c_1c_2^{-1} & I
    \end{array}
  \right)
\end{eqnarray*}
Although in principal this requires transforming $\boldsymbol{\psi}_{\parallel},\boldsymbol{\psi}_{\perp}$ to
new coordinates, in practice, $c_0 =\Or(r^4), c_1=\Or(r^2), c_2=\Or(1)$  so that to fourth order in the $r$ expansion of $\mathcal{C}(r)$, $u$ can be taken as the identity matrix.

Substituting these expressions  for the vectors and matrices into Eq.~(\ref{eq:Correlation3}),
performing the inverse Fourier transform, and rescaling the coordinates as:
\begin{eqnarray*}
\boldsymbol{\psi}_{\parallel}\to r^{2}\boldsymbol{\psi}_{\parallel} & , &~ c_3 \to r^{-4}c_3\\
\boldsymbol{\psi}_{-}\to r\boldsymbol{\psi}_{-} & , &~ \Delta b \to r^{-2}\Delta b
\end{eqnarray*}
converts $\tilde{B}$ into a block-diagonal form, and thus
\begin{equation*}
  \boldsymbol{\psi}\tilde{B}^{-1}\boldsymbol{\psi} \to
  \boldsymbol{\psi}_{+}
  \left(
    \begin{array}{cc}
      c_3 & 0 \\
      0 & c_2
    \end{array}
  \right)^{-1} \boldsymbol{\psi}_{+}
  +
  \boldsymbol{\psi}_{-} \Delta b^{-1} \boldsymbol{\psi}_{-}
\end{equation*}
This form is convenient for expansion in $r$, yet in order to evaluate the integral Eq. ~(\ref{eq:Correlation3}) one should also
expand $\tilde{\rho}$ as a power series in $r$. Here care must be taken, as $\tilde{\rho}$ is not a smooth function. Let us define a shorthand notation:
\begin{equation*}
  h(\boldsymbol{\psi}_i) = \det H[\phi(\mathbf{r}_i)].
\end{equation*}
so that
\begin{equation*}
  \tilde{\rho}(0)\tilde{\rho}(r) = h(\boldsymbol{\psi}_1)h(\boldsymbol{\psi}_2) Q(\boldsymbol{\psi}_1,\boldsymbol{\psi}_2)
\end{equation*}
where $Q$ is some function which depends on the type of the density we are dealing with (unsigned or minima). Then we can expand the Hessian determinants as
\begin{eqnarray*}
\fl h\left(\boldsymbol{\psi}_1\right)h\left(\boldsymbol{\psi}_2\right) & = & h\left(\frac{\boldsymbol{\psi}_{\perp}+r^{2}\boldsymbol{\psi}_{\parallel}+r\boldsymbol{\psi}_{-}}{\sqrt{2}}\right)h\left(\frac{\boldsymbol{\psi}_{\perp}+r^{2}\boldsymbol{\psi}_{\parallel}-r\boldsymbol{\psi}_{-}}{\sqrt{2}}\right)\\
\fl & = & h^{2}\left(\frac{\boldsymbol{\psi}_{\perp}}{\sqrt{2}}\right)+\frac{r^{2}}{2}\boldsymbol{\psi}_{-}\cdot\left[h\boldsymbol{\nabla} \otimes \boldsymbol{\nabla} h-\boldsymbol{\nabla} h\otimes\boldsymbol{\nabla} h \right]\left(\frac{\boldsymbol{\psi}_{\perp}}{\sqrt{2}}\right)\cdot\boldsymbol{\psi}_{-}+\Or\left(r^{4}\right)
\end{eqnarray*}
Now, recall that the Hessian matrix one can construct from the elements of $\boldsymbol{\psi}_{\perp}$ is
\begin{eqnarray}
\left(\begin{array}{ccc}
0 & 0 & 0\\
0 & \Phi_{22} & \Phi_{23}\\
0 & \Phi_{23} & \Phi_{33}
\end{array}\right).
 \end{eqnarray}
 This implies that  $h\left(\frac{\boldsymbol{\psi}_{\perp}}{\sqrt 2}\right)=0$ since both the first row and first column equal zero. In addition, of all the elements of the vector $\nabla h$ only the first element survives:
\begin{eqnarray}
\frac{\partial}{\partial_{\Phi_{11}}}\det{H}=\det H_{\perp}=\det\left(\begin{array}{cc}
\Phi_{22} & \Phi_{23}\\
\Phi_{23} & \Phi_{33}
\end{array}\right)
\end{eqnarray}.
This formula holds for any dimension. It allows us to write for the unsigned correlation:

\begin{eqnarray}\label{eq:final-short-uns}
  \fl \left\langle
    \tilde{\rho}\left(0\right)
    \tilde{\rho}\left(r\right)\right\rangle  & = & \left\langle \left|-\frac{r^{2}}{2}\boldsymbol{\psi}_{-}\left(\boldsymbol{\nabla}h \otimes\boldsymbol{\nabla}h\right) \boldsymbol{\psi}_{-}+\Or\left(r^{4}\right)\right|\right\rangle \nonumber \\
\fl & = & \frac{r^{2}}{2} \Delta b_{11}\left\langle \left(\det H_{\perp}\right)^{2}\right\rangle+\Or\left(r^{4}\right)
\end{eqnarray}
thus recovering our main result, Eq. (\ref{eq:main-result}), with:
\begin{equation}
  \label{eq:alpha3}
  \alpha_3 = \frac{3^{\frac{d-1}{2}}}{\left(2\pi\sqrt{g_2 g_4}r\right)^{d}}\cdot  \Delta b_{11}\left\langle \left(\det H_{\perp}\right)^{2}\right\rangle
\end{equation}
Note that $ \left\langle \left(\det H_{\perp}\right)^{2}\right\rangle$ is the expectation of a polynomial in gaussian variables and can be evaluated with Wick's theorem.

For the minima-minima correlation we find:
\begin{eqnarray*}
\fl \left\langle \left[\tilde{\rho}\left(0\right)\tilde{\rho}\left(r\right)\right]\right\rangle  & = & \left\langle \left[-\frac{r^{2}}{2}\boldsymbol{\psi}_{-}\left(\boldsymbol{\nabla}h \otimes\boldsymbol{\nabla} h\right) \boldsymbol{\psi}_{-}+\Or\left(r^{4}\right)\right]\right.\\
  &\times & \left. \Theta\left[-\frac{r^{2}}{2}\boldsymbol{\psi}_{-}(\boldsymbol{\nabla}h\otimes\boldsymbol{\nabla}h)\boldsymbol{\psi}_{-}+\Or\left(r^{4}\right)\right] \prod_{i=0}^{d-2}\Theta\left(\det{H_{\perp}}^{\left(i\right)}+\Or\left(r^2\right)\right)\right\rangle\nonumber\\
\fl & = & \Or\left(r^{4}\right)\end{eqnarray*}
To obtain this result we have used the fact that $\boldsymbol{\psi}_{-}(\boldsymbol{\nabla}h \otimes\boldsymbol{\nabla}h) \boldsymbol{\psi}_{-}$ is negative semidefinite. Also, note the fact that all the minors of ${H}$ at both points can be approximated by ${H_{\perp}}$ so that the two sets of $\Theta$-functions (from the two densities) merge.

Obtaining the next order is considerably harder. However, we can deduce its
power law dependence from the following considerations. Let us concentrate
on the minima-minima correlation, so that:
\[
\left\langle \left[\tilde{\rho}\left(0\right)\tilde{\rho}\left(r\right)\right]\right\rangle \sim r^{2}\left\langle \left[-\chi^{2}h_{\perp}^{2}+\mu r^{2}\right]\Theta\left[-\chi^{2}h_{\perp}^{2}+\mu r^{2}\right]\right\rangle
\]
Here $\chi$ is the first element of $\boldsymbol{\psi}_{-}$, $h_{\perp}=\det H_{\perp}$, and $\mu$
is some function of $\boldsymbol{\psi}_{\perp}$, $\boldsymbol{\psi}_{\parallel}$, and $\boldsymbol{\psi}_{-}$.
Now, for the $\Theta$ function to be nonzero we require
\begin{equation}
  \label{eq:theta-condition}
  \frac{\chi^{2}\eta_{\perp}^{2}}{\mu}< r^{2}\ll1.
\end{equation}

We can essentially ignore $\mu$ in this equation since as long
as $\mu$ is not of order $\sim1$ this condition doesn't hold,
so we have $\chi^{2}\eta_{\perp}^{2}\ll r^{2}\ll1$. Let us change
coordinates by replacing $\chi$ with $x=\chi\eta_{\perp}$,
keeping all other variables. This gives:

\begin{eqnarray}
\left\langle \left[\tilde{\rho}\left(0\right)\tilde{\rho}\left(r\right)\right]\right\rangle  & \sim & r^{2}\int dx\left\langle \left[-x^{2}+\mu r^{2}\right]\Theta\left[r^{2}-x^2\right]\right\rangle \nonumber \\
 & = & r^{2}\left.\left(-\frac{x^{3}}{3}+\mu r^{2}x\right)\right|_{-r}^{r} \propto r^{5} \label{eq:final-short-min}
\end{eqnarray}
and therefore ${\cal C}\left(r\right)  \propto   r^{5-d}$ which is the result stated in Eq. (\ref{eq:main-result}) for $k=3$.
For the unsigned correlation, there is no condition such as (\ref{eq:theta-condition}), so the integration will simply give a correction of order $r^{2}$ above the first order. This can be seen by noting
 the identity $\left|x\right|=2x~\Theta\left(x\right)-x$.
\noindent

\subsection*{Application of short order analysis in two and three dimensions}
\label{sec:appl-short-append}

Let us apply the short-distance result expressed in Eq. (\ref{eq:final-short-uns}) to the two and three dimensional problem.
To this end we apply Wick's theorem to Eq.~(\ref{eq:alpha3}) and find that, in two dimensions, the unsigned correlation behaves as
\begin{equation}
{\cal C}\left(r\right) = \alpha_3 + \Or\left(r^2\right)\label{eq:CunsShort2d}
\end{equation}
with
\begin{eqnarray}
\alpha_3 = \frac{1}{6\sqrt{3}\pi^{2}} \left(k_{1}^2 -
k_{0}^2\right),
\end{eqnarray}
where $k_{0}=g_4 / g_2$, and $k_{1}^2 = g_6 / g_2$, while for the minima-minima correlation we have ${\cal C}\left(r\right) \propto r^3 $.

In the three dimensional case the short range asymptotics of the unsigned correlation function takes the form
\begin{equation}
  \mathcal{C}(r) = \alpha_3  \frac{1}{r} + \Or\left(r\right) \label{eq:CunsShort3d}
\end{equation}
with
\begin{eqnarray}
\alpha_3 = \frac{29}{288\pi^{3}} \sqrt{k_0}\left(k_1^2 - k_0^2\right),
\end{eqnarray}
while for the minima-minima correlation we find $\mathcal{C}(r) \propto r^2$.

\begin{figure}
\subfloat[\label{fig:short2dMin} Minima-minima correlation in 2D.]{\includegraphics[clip,width=0.45\textwidth]{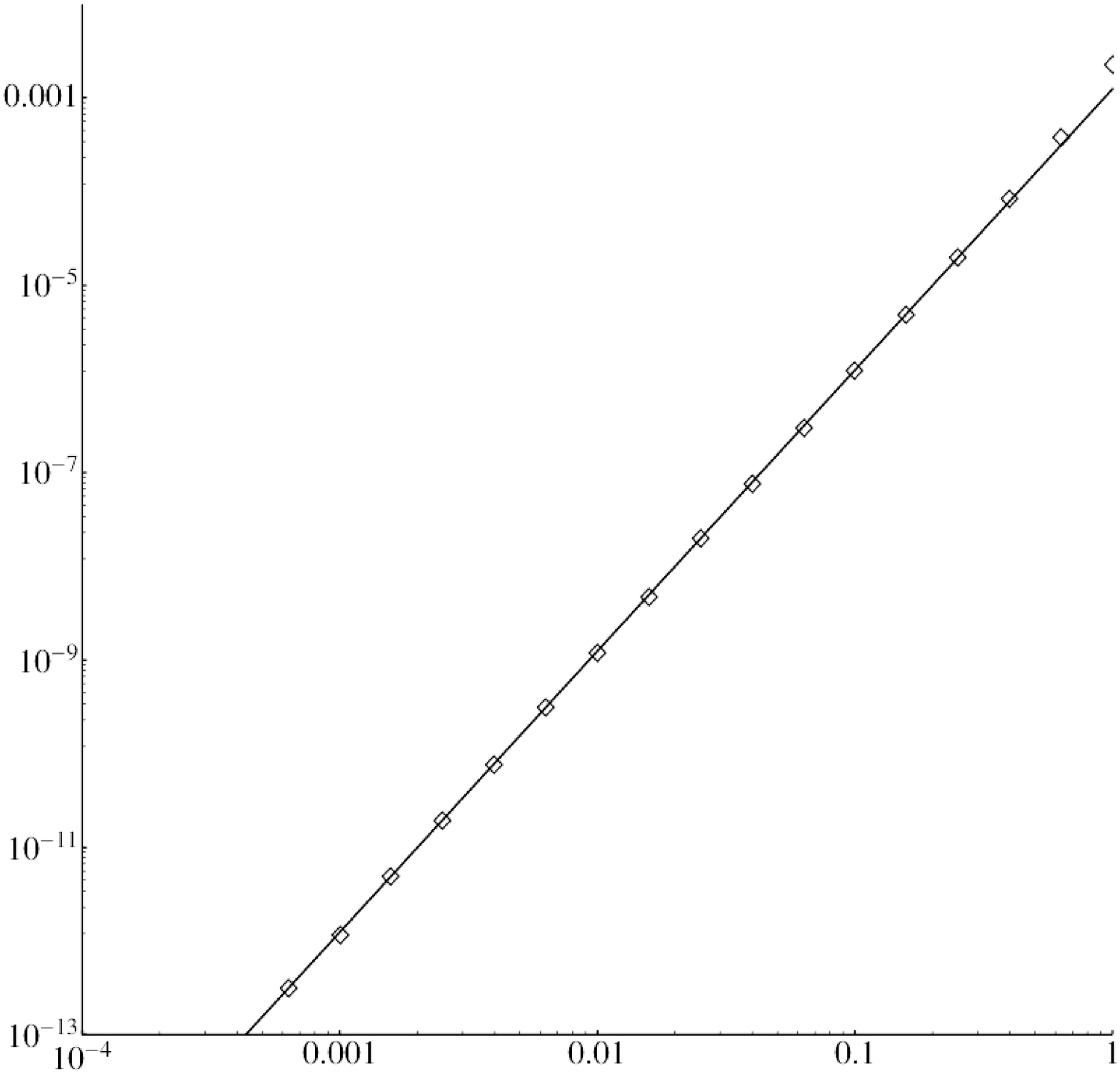}

}\hfill{}\subfloat[{\label{fig:short2dAbs}Unsigned correlation in 2D.}]{\includegraphics[clip,width=0.45\textwidth]{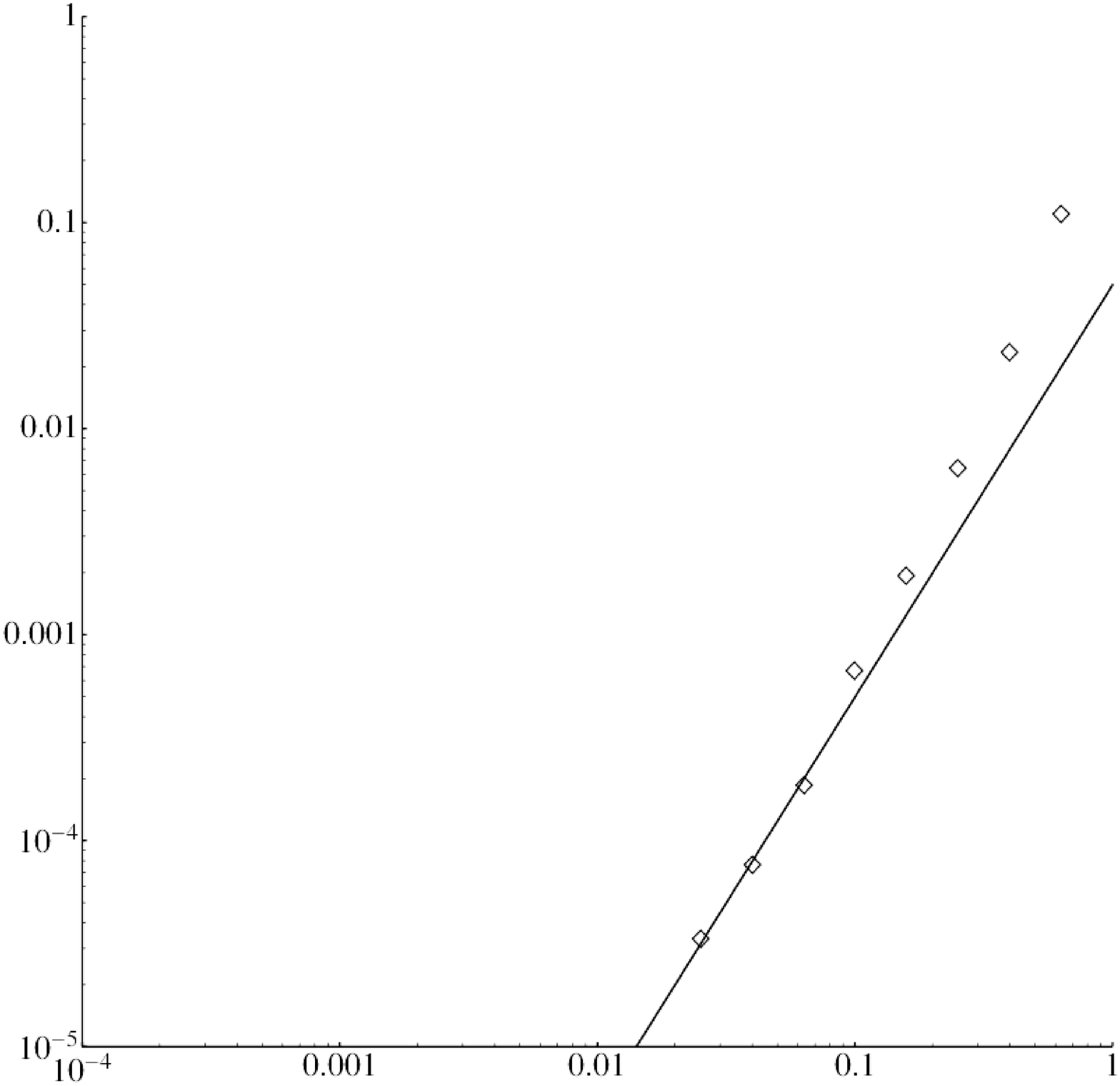}

}

\subfloat[\label{fig:short3dMin}Minima-minima correlation in 3D.]{\includegraphics[clip,width=0.45\textwidth]{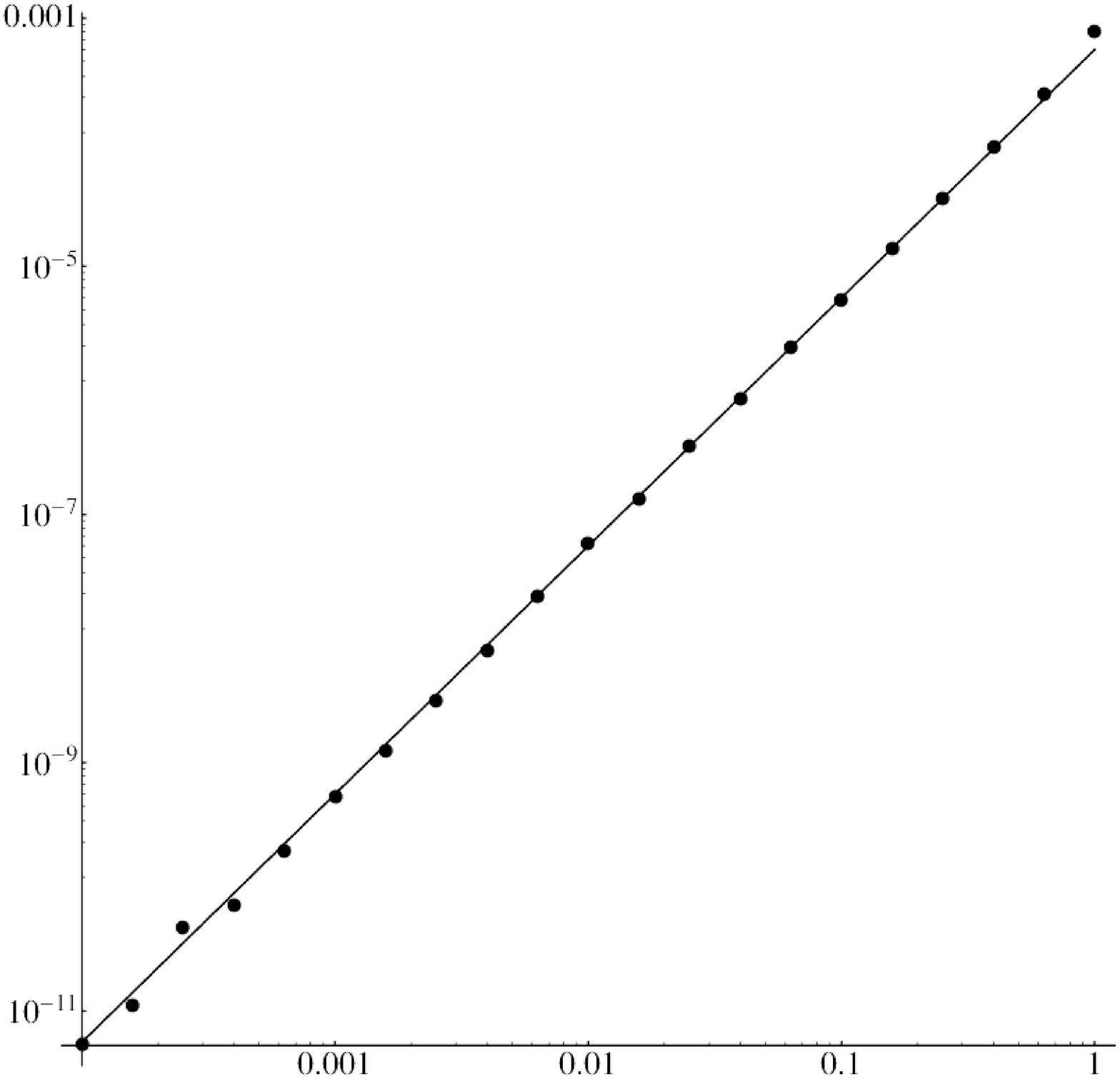}

}\hfill{}\subfloat[{\label{fig:short3dAbs}Unsigned correlation in 3D.}]{\includegraphics[clip,width=0.45\textwidth]{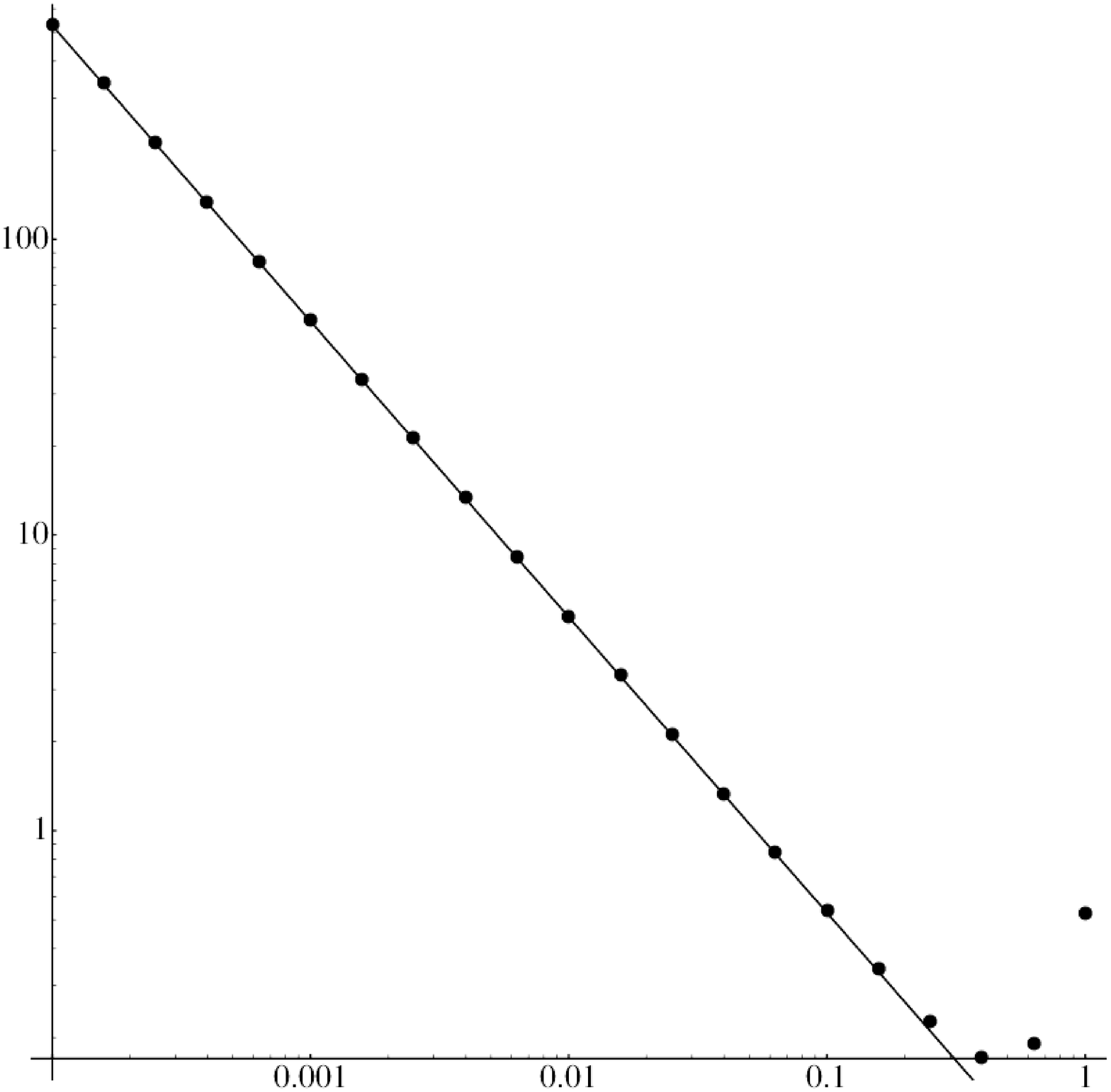}

}

\caption{The short range limit of correlation functions of critical points in two and three dimensions.
Solid lines represent the analytic result expressed in Eq.~ (\ref{eq:main-result}) while dots represent the 
result of a numerical calculation (see Appendix B) which has been performed for a gaussian random field with 
correlations defined by Eq.~(\ref{eq:Gd}).}\label{fig:Short-range-correlation}
\end{figure}

To verify our analysis, we compare our results to numerical calculations. The correlation functions chosen are the same as those appearing in sections \ref{sec:app-two-dimensions} and \ref{sec:three-dimensions}. Figure \ref{fig:Short-range-correlation} shows the excellent agreement we obtained with numerical calculations. There is one problematic graph, namely \ref{fig:short2dAbs}. The reason for this is simply the technical difficulty in numerically extracting the next-leading order behavior for this problem.

\section{Details of numerical analysis}
\label{sec:deta-numer-analys}

In this section we outline the methods we used to evaluate correlations numerically. From Eq. (\ref{eq:Correlation3}) in Sec. \ref{sub:Presentation-of-the} it follows that one can represent the correlation in the form:

\begin{eqnarray}
\fl {\cal C}(r) = \frac{1}{(2\pi)^{d}\sqrt{\det A}}\cdot \frac{1}{(2\pi)^{2(n-d)}\sqrt{\det \tilde{B}}} \int d\psi \tilde{\rho} (0) \tilde{\rho(r)} \exp \left( -\frac{1}{2} \boldsymbol{\psi} \tilde{B}^{-1} \boldsymbol{\psi}\right)
\end{eqnarray}
Here the elements of both $A$ and $\tilde{B}$ are functions of $r$. Evaluating the $\det A$ term numerically poses no difficulty. In order to evaluate the integral, we perform a Cholesky decomposition \cite{Press2007} numerically on $\tilde{B}$:
\begin{eqnarray}
  \label{eq:cholesky}
  \tilde{B} = V^t V \\
  \boldsymbol{\psi} \to V^{-1} \boldsymbol{\psi}
\end{eqnarray}
This transforms the integral to an integral over an appropriate set of standard normal variables: $\boldsymbol{\psi}_i \sim N(0,1)$. The integral can then be estimated by standard Monte-Carlo integration. Our integrations were performed with Sobol numbers of the appropriate dimension generated by Mathematica 7.

The evaluations of the two-dimensional problem required $2^{23}$ lattice points for every data point appearing in figure
 \ref{fig:2dFar} and $2^{24}$ lattice points for every data point in figure \ref{fig:Short-range-correlation}. For the three dimensional problem, we used $2^{25}$ lattice points for data points in figure \ref{fig:3dFar}, and between $0.875$ and  $ 1.125 \times 2^{26}$ points (depending on the specific dataset) for data points in figure \ref{fig:Short-range-correlation}.

\bibliographystyle{unsrt}
\bibliography{correlation}

\end{document}